\documentstyle[11pt,paspconf,epsf]{article}

\begin{document}

\title{The evolution of groups and clusters }
\author{S. Gottl\"ober}  
\affil{Astrophysical Institute Potsdam, An der Sternwarte 16, 14482
Potsdam, Germany}
\author{A. Klypin and. A. V. Kravtsov}  
\affil{Astronomy Department, NMSU, Dept.4500, Las Cruces, NM
88003-0001, USA}
\author{V. Turchaninov}  
\affil{Keldysh Institute for Applied Mathematics,
Miusskaja Ploscad 4,  125047 Moscow, Russia}

\begin{abstract}
Using high resolution $N$-body simulations we study the
formation and evolution of clusters and groups in a
$\Lambda$CDM cosmological model. Groups of galaxies have been formed
already before $z=4$. Merging of groups and accretion leads to cluster
formation at $z{_ <\atop{^\sim}} 2$. Some of the groups merge into
large isolated halos.
\end{abstract}


\section{Introduction}

Cosmological scenarios with cold dark matter alone cannot explain the
structure formation both on small and very large scales.  Scenarios
with a non-zero cosmological constant $\Lambda$ have been proved to be
very successful in describing most of the observational data at both
low and high redshifts. Moreover, from a recent analysis of 42
high-redshift supernovae Perlmutter et.al (1999) found direct evidence
for $\Omega_{\Lambda}= 0.72$ within a flat cosmology. 

It is generally believed that the dark matter component of galaxies,
the dark matter halos, plays an important role in the formation of
galaxies. Properties and evolution of the halos depend on the
environment (Avila-Reese et al. 1999, Gottl\"ober et al. 1999a,b) which 
implies, in turn, that properties of galaxies can be expected to depend on 
cosmological environment too. Many of the dark matter halos are much more
extended than galaxies and contain several subhalos. Here we present results of a
study of the formation and evolution of these large halos which host
groups and clusters of galaxies.

\section {Dark matter halos in the simulation}

We simulate evolution of $256^3$ cold dark matter particles in a
$\Lambda$CDM model ($\Omega_M=1-\Omega_{\Lambda}=0.3$; $\sigma_8=1.0$;
$H_0=70$ km/s/Mpc). As a compromise we have
chosen a simulation box of $60 h^{-1} {\rm Mpc}$ in order to study the
statistical properties of halos in a cosmological environment and to
have also a sufficient mass resolution (particle mass of $1.1 \times
10^9 h^{-1} {\rm M_{\odot}}$). The simulations were done using the
Adaptive Refinement Tree (ART) code (Kravtsov, Klypin \& Khokhlov
1997). The code used a $512^3$ homogeneous grid on the lowest level of
resolution and six levels of refinement, each successive refinement
level doubling the resolution. The sixth refinement level corresponds
to a formal dynamical range of 32,000 in high density regions. Thus we
can reach in a 60 $h^{-1}$ Mpc box the force resolution of $\approx
2h^{-1}$ kpc.

Identification of halos in dense environments and reconstruction of
their evolution is a challenge. In order to find halos in the
simulation we have developed two algorithms, which we called the {\em
hierarchical friends-of-friends} (HFOF) and the {\em bound density
maxima} (BDM) algorithms (Klypin et al. 1999). These algorithms work on
a snapshot of the particle distribution. They are able to identify
``halos inside of halos'', i.e. stable gravitationally self-bound halos
which move inside of a larger region of virial overdensity.  Therefore,
we can find dark matter halos which correspond to galaxies inside of
groups or clusters but also small satellites bound to a larger galaxy
(like the LMC and the Milky Way or the M51 system). The algorithm
identifies halos at different redshifts. In a second step we establish
the ``ancestor-descendant'' relationships for all halos at all times
(Gottl\"ober et al. 1999a,b).

The total mass of a halo depends on its assumed radius. However, for
satellite halos inside larger bound systems the notion of radius
becomes somewhat arbitrary.  The formal virial radius of a satellite
halo is simply equal to the virial radius of its host halo. We have
chosen to define the outer {\em tidal} radius of the satellite halos as
the radius at which their density profile starts to flatten. We try to
avoid the problem of mass determination by assigning not only a mass to
a halo, but finding also its maximum ``circular velocity''
($\sqrt{GM/R}$), $v_{circ}$. This is the quantity which is more
meaningful observationally. Numerically, $v_{circ}$ can be measured
more easily and more accurately then mass.

\begin{figure}[t]
\centering 
\leavevmode
\epsfxsize=0.7\textwidth
\epsfbox{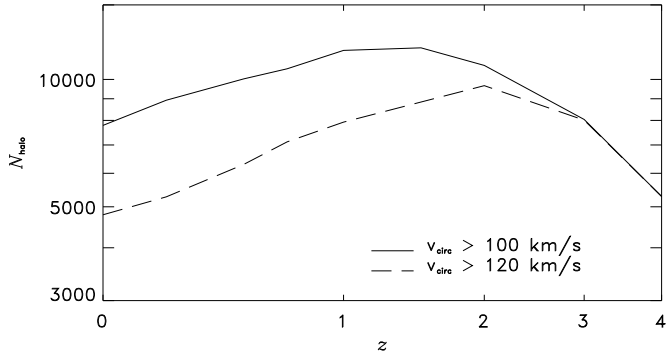}
\caption[]{The number of halos with more than 50 bound particles 
depending on redshift for $v_{circ} > 100$km/s (solid line) 
and $v_{circ} > 120$km/s (dashed line)}
\label{all}
\end{figure}

In order to define a complete halo sample that is not affected by the
numerical details of halo finding procedure we have constructed the
differential velocity functions at $z=0$ for different mass thresholds
and maximum radii. We found that the halo samples do not depend on the
numerical parameters of the halo finder for halos with $v_{circ} {_
>\atop{^\sim}} 100$ km/s (Gottl\"ober et al. 1999a). Here we want to
study samples of halos at redshifts $z \le 4$ with circular velocities
$v_{circ} > 100$ km/s and, for comparison, samples of more massive
halos with circular velocities $v_{circ} > 120$ km/s.  Moreover, we
include only halos which consist of at least 50 bound particles. This
reduces the possible detection of fake halos in the simulation, in
particular the detection of very small fake satellite halos.  At
redshift $z=0$ we have detected 7786 halos with $v_{circ} > 100$km/s in
our simulation, which corresponds a number densities of $0.036 h^3 {\rm
Mpc}^{-3}$. 4787 of these halos have circular velocities $v_{circ} >
120$km/s ($0.022 h^3 {\rm Mpc}^{-3}$). These densities roughly agree
with the number density of galaxies with magnitudes $M_r {_
<\atop{^\sim}} -16.3$ and $M_r {_ <\atop{^\sim}} -18.4$ estimated by
the luminosity function of the Las Campanas redshift survey (Lin et
al. 1996) which we have extended to -16.3.

The time evolution of the total number of halos in the box is shown in
Fig.~{\ref{all}}. At $z=4$ we found the same number of halos
independent of the threshold of the circular velocity. This means that
we have reached here the resolution limit of the simulation. In fact,
we had excluded low mass halos with less than 50 particles.

\section {Environment of Halos}

To find the cosmological environment of each of these halos at $z \le
4$ we run a friend-of-friend analysis over the dark matter particles
with a linking length of 0.2 times the mean interparticle distance. By
this procedure we find clusters of dark matter particles with an
overdensity of about 200. At $z=0$ the virial overdensity in the
$\Lambda$CDM model under consideration is about 330 which corresponds
to a linking length of about 0.17. With increasing $z$ the virial
overdensity decreases and reaches 200 at $z=1$. Here we have used the
same linking length for all $z$. Therefore, the objects which we find
at $z<1$ are slightly larger than the objects at virial overdensity. We
have increased the linking length at $z=0$ because we found in the
vicinity of galaxy clusters halos which had already interacted with the
cluster but were in the moment of detection just outside the region of
virial overdensity.

In a second step we find for each of the halos at all $z$ the cluster
of dark matter particles to which the halo belongs.  We call the halo a
cluster galaxy if the the halo belongs to a particle cluster with a
total mass larger than $10^{14} h^{-1} {\rm M_{\odot}}$. We call it an
isolated galaxy if only one halo belongs to the object at overdensity
200 which we found by the friends-of-friends algorithm. This definition
ensures that the ``isolated galaxies'' really do not interact with
other galaxies. We found a substantial number (about 10 \%) of
doublets. In some cases we found pairs of halos with approximately the
same mass but in most of the cases these doublets consist of one big
halo with a small bound satellite which is inside of or partly overlaps
with the larger halo. On the one hand, there could be more small
satellites which we do not detect due to the limited mass resolution so
that the doublet is in fact a small group. On the other hand, there
could be really only one small satellite so that the object better
could be handled as an isolated galaxy. In order to avoid difficult
decisions, we have handled the galaxy pairs separately having in mind
that some of them could belong to isolated galaxies whereas other are
seeds of a small group.  The rest of halos are called group galaxies.

The 7786 halos with $v_{circ} > 100$km/s are distributed over 18
clusters, 252 groups and 373 pairs, 4892 halos are isolated.
  
\section {The Evolution of Halos }

\begin{figure}[t]    
\centering 
\leavevmode
\epsfxsize=0.7\textwidth
\epsfbox{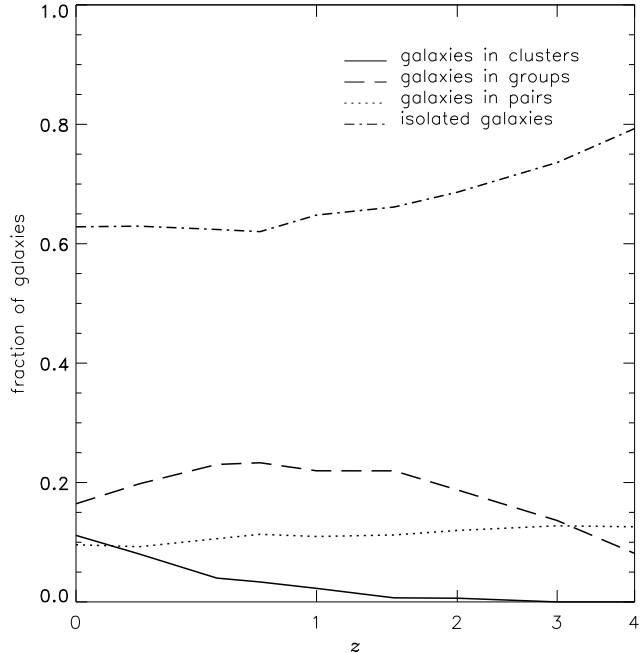}
\caption[]{The evolution of the fraction of galaxies in clusters,
groups and pairs and of isolated galaxies for dark matter halos with
$v_{circ} > 100$km/s}
\label{frac100}
\end{figure}

The total number of halos with $v_{circ} > 100$ km/s increases from
5290 at $z=4$ to the maximum value of 11847 at redshift $z=1.5$ and
then decreased to 7786 at redshift $z=0$ (see Fig.~1).  Assuming that
each of these halos would contain a galaxy we compute the evolution of
the fraction of isolated ``galaxies'' and ``galaxies'' in clusters,
groups, and pairs.

The first object of virial mass $>10^{14} h^{-1} {\rm M_{\odot}}$ forms
between $z=2.5$ and $z=2$. At $z=2$ this cluster of $1.3 \times 10^{14}
h^{-1} {\rm M_{\odot}}$ contains already 68 satellite halos.  The most
massive central halo of this cluster has a circular velocity of
760~km/s. The number of cluster galaxies increases up to $z=0$, where
868 galaxies have been detected in clusters.

At all analyzed time epochs in the simulation, we found approximately
10~\% of pairs. We have detected 113 groups already
at $z=4$. These groups contain 430 galaxies. The number of group
galaxies increases very rapidly and reaches a maximum of 2607 at
$z=1.5$. Afterwards it decreases up to 1280 at $z=0$. Finally, also the
number of isolated galaxies increases from $z=4$ (4194) until $z=1.5$
(7854) and then decreases until $z=0$ (4892).

In Fig. \ref{frac100} we show the time evolution of the fraction of
isolated galaxies and galaxies in clusters, groups, and pairs for dark
matter halos with $v_{circ} > 100$km/s. The sample of more massive
halos ($v_{circ} > 120$km/s) shows essentially the same behavior, but
the total number of halos is smaller.

\begin{figure}[t]    
\centering 
\leavevmode
\epsfxsize=0.7\textwidth
\epsfbox{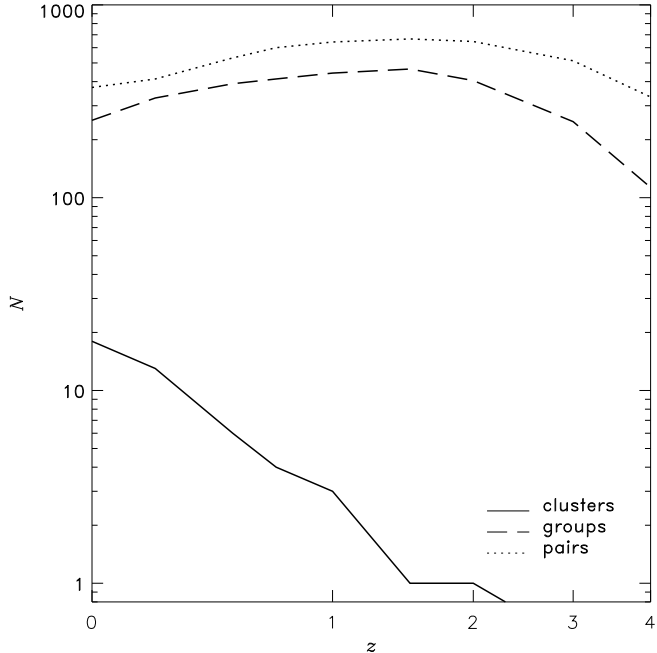}
\caption[]{The evolution of the number of clusters,
groups and pairs for dark matter halos with $v_{circ} > 100$km/s}
\label{numbers}
\end{figure}

Now let us discuss the evolution of clusters and groups. To this end,
we find similar to the chain of progenitors of each halo the
progenitors of the clusters and groups. The first cluster has been
formed before $z = 2$. Due to accretion and merging of groups the
number of clusters increases to 18 at $z=0$ (Fig. \ref{numbers}). At
$z=4$ we found 113 groups and 333 pairs in our simulation. It is
remarkable that these numbers do not depend on the chosen threshold of
the circular velocity. Both the number of groups and pairs increases up
to $z=1.5$ where we have 465 (363) groups and 666 (567) pairs for
$v_{circ} > 100$km/s ($v_{circ} > 120$km/s). At $z=0$, 252 (152) groups
and 373 (258) pairs remain. The total number of groups and pairs
reduces with increasing threshold of the circular velocity. In fact, if
we take into account only more massive objects some of the pairs would
be identified as isolated galaxies whereas some of the groups would be
identified as pairs or isolated galaxies. However, as we already
mentioned, the overall fraction of galaxies in groups and pairs is
rather insensitive to this threshold.

From Fig. \ref{frac100}, we see that the fraction of galaxies in groups
decreases from about 22\% at $z=0.7$ to about 16\% at $z=0$.  At the
same time, the fraction of isolated halos slightly increases. The
decreasing number of groups is mostly due to the fact that groups merge
into more massive objects. However, we also found a small fraction of
groups which merge into isolated halos. In the simulation we found 20
halos with masses larger than $10^{12} h^{-1} {\rm M_{\odot}}$ the
progenitor of which at $z=1$ was a group with 4 to seven members. Such
a merged group could appear today as an isolated elliptical with a
group-like X-ray halo (Mulchaey \& Zabludoff 1999; Vikhlinin et
al. 1999).

\section {Conclusions}

The general trend during evolution of clustering is the formation of
small bound systems of halos at $z{_ >\atop{^\sim}} 1.5$. After $z=1.5$
these systems tend to merge and to increase mass by accretion so that
large galaxy clusters grow in the simulation.  The total number of
small bound systems and the total number of galaxies in these small
systems rapidly decreases after $z=1.5$. The fraction of isolated
galaxies remains approximately constant after $z =1$, whereas the
fraction of galaxies in groups decreases. Some of the groups merge
after $z=1$ to form large isolated galaxies.

\acknowledgments This work was funded by the NSF and NASA grants to
NMSU.  SG acknowledges support from Deutsche Akademie der
Naturforscher Leopoldina with means of the Bundesministerium f\"ur
Bildung und Forschung grant LPD 1996.  We acknowledge support by NATO
grant CRG 972148.

\end{document}